\documentclass[twocolumn]{aastex631}

\usepackage{soul}
\usepackage{multirow}
\usepackage{epstopdf}
\AppendGraphicsExtensions{.tif}
\usepackage[hang,flushmargin]{footmisc}
\usepackage{natbib}

\shorttitle{Studying Crab X-ray Polarization Using Deeper IXPE Observations}
\shortauthors{Wong et al.}

\begin{document}

\title{Analysis of Crab X-ray Polarization using Deeper IXPE Observations}

\correspondingauthor{Josephine Wong}
\email{joswong@stanford.edu}
\author[0000-0001-6395-2066]{Josephine Wong}
\affiliation{Department of Physics and Kavli Institute for Particle Astrophysics and Cosmology, Stanford University, Stanford, California 94305, USA}

\author[0000-0001-7263-0296]{Tsunefumi Mizuno}
\affiliation{Hiroshima Astrophysical Science Center, Hiroshima University, 1-3-1 Kagamiyama, Higashi-Hiroshima, Hiroshima 739-8526, Japan}

\author[0000-0002-8848-1392]{Niccol\'{o} Bucciantini}
\affiliation{INAF Osservatorio Astrofisico di Arcetri, Largo Enrico Fermi 5, 50125 Firenze, Italy}
\affiliation{Dipartimento di Fisica e Astronomia, Universit\`{a} degli Studi di Firenze, Via Sansone 1, 50019 Sesto Fiorentino (FI), Italy}
\affiliation{Istituto Nazionale di Fisica Nucleare, Sezione di Firenze, Via Sansone 1, 50019 Sesto Fiorentino (FI), Italy}

\author[0000-0001-6711-3286]{Roger W. Romani}
\affiliation{Department of Physics and Kavli Institute for Particle Astrophysics and Cosmology, Stanford University, Stanford, California 94305, USA}

\author[0000-0001-9108-573X]{Yi-Jung Yang}
\affiliation{Graduate Institute of Astronomy, National Central University, 300 Zhongda Road, Zhongli, Taoyuan 32001, Taiwan}
\affiliation{Laboratory for Space Research, The University of Hong Kong, Cyberport 4, Hong Kong}

\author[0009-0007-8686-9012]{Kuan Liu}
\affiliation{Guangxi Key Laboratory for Relativistic Astrophysics, School of Physical Science and Technology, Guangxi University, Nanning 530004, China}

\author[0000-0002-9370-4079]{Wei Deng}
\affiliation{Guangxi Key Laboratory for Relativistic Astrophysics, School of Physical Science and Technology, Guangxi University, Nanning 530004, China}

\author{Kazuho Goya}
\affiliation{Hiroshima University, School of Science, 1-3-1 Kagamiyama, Higashi-Hiroshima, Japan}

\author[0000-0002-0105-5826]{Fei Xie}
\affiliation{Guangxi Key Laboratory for Relativistic Astrophysics, School of Physical Science and Technology, Guangxi University, Nanning 530004, China}
\affiliation{INAF Istituto di Astrofisica e Planetologia Spaziali, Via del Fosso del Cavaliere 100, 00133 Roma, Italy}

\author[0000-0001-7397-8091]{Maura Pilia}
\affiliation{INAF Osservatorio Astronomico di Cagliari, Via della Scienza 5, 09047 Selargius (CA), Italy}

% ...

\begin{abstract}

We present Crab X-ray polarization measurements using IXPE data with a total exposure of 300ks, three times more than the initial 2022 discovery paper. Polarization is detected in three times more pulsar phase bins, revealing an S-shaped $+40^\circ$ polarization angle sweep in the main pulse and ${>}1\sigma$ departures from the OPTIMA optical polarization in both pulses, suggesting different radiation mechanisms or sites for the polarized emission at the two wavebands. Our polarization map of the inner nebula reveals a toroidal magnetic field, as seen in prior IXPE analyses. Along the southern jet, the magnetic field orientation relative to the jet axis changes from perpendicular to parallel and the polarization degree decreases by ${\sim}6\%$. These observations may be explained by kink instabilities along the jet or a collision with a dense, jet-deflecting medium at the tip. Using spectropolarimetric analysis, we find asymmetric polarization in the four quadrants of the inner nebula, as expected for a toroidal field geometry, and a spatial correlation between polarization degree and photon index.

\end{abstract}



\section{Introduction} \label{sec:intro}

\begin{figure*}
    \centering
    \includegraphics[width=\linewidth]{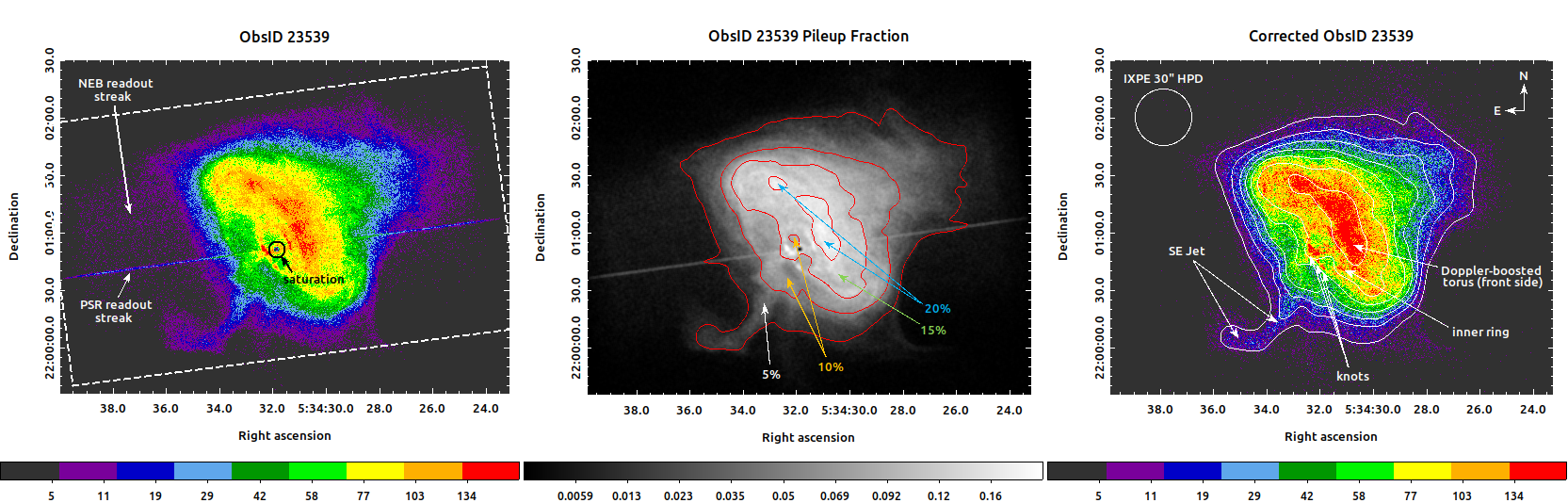}
    \caption{Chandra ObsID 23539 of the Crab Nebula before (left) and after (right) processing, displaying a $3.9'\times2.9'$ region. Images are matched in log scale. CCD readout streaks and photon pileup at the pulsar position and throughout the nebula in the original Level 2 events file were corrected. Center image shows map with red iso-contours of the pileup fraction in the nebula, which reaches up to $\sim 20\%$ in the Doppler-boosted region. See Section \ref{subsec:cxo_obs} for a discussion of the processing technique.}
    \label{fig:cxo}
\end{figure*}

\begin{table*}
\centering
\begin{tabular}{cccc}
\hline
Parameters & Ephemeris 1 & Ephemeris 2 & Ephemeris 3 \\
\hline
JBO Month  & March 2022 & February 2023 & September 2023 \\
PEPOCH (MJD) & 59625 & 59990 & 60202 \\
$\nu$ (Hz) & 29.5870753202 & 29.5754926637 & 29.5687705364 \\
$\dot\nu\:(10^{-10}\:\mathrm{Hz} \cdot \mathrm{s}^{-1})$ & -3.67470
& -3.67089 & -3.66897 \\
$\ddot\nu\:(10^{-21}\:\mathrm{Hz} \cdot \mathrm{s}^{-2})$ & 8.04 & 5.24 & 11.6 \\
\hline
\end{tabular}
\caption{Jodrell Bank Observatory (JBO) ephemerides used in the phase-folding of the IXPE Crab observations. Ephemeris 1 was applied to ObsID 01001099, Ephemeris 2 to ObsID 02001099, and Ephemeris 3 to ObsID 02006001.}
\label{tab:ephemeris}
\end{table*}

Pulsar wind nebulae (PWNe) are highly-energetic astrophysical sources that consist of a central spinning neutron star (pulsar) whose powerful magnetic field ($B\sim10^{12}\ \mathrm{G}$) generates a wind of relativistic ($\gamma \gtrsim 10^5$) electrons and positrons that escape along open field lines, impinging on and carrying energy into the surrounding SN ejecta or ISM \cite{Kirk2009}. They can be detected across the entire electromagnetic spectrum with a non-thermal spectral energy distribution (SED) that displays a synchrotron bump that extends from the radio to hard X-rays (up to MeV) and an inverse Compton bump that can reach up to TeV \citep{Slane2017} or even PeV \citep{Liu2021} energies. Spatially-resolved observations reveal time-varying structures such as wisps, knots, filaments, and polar jets that point to an ongoing resupply of energy whose source we now know to be the pulsar. 

The pulsar itself is seen as a bright point source at the center of the PWN and its radiative signature is the pulsed light curve, which exhibits a consistent double-peaked profile from radio to gamma-rays. The origin of this radiation is believed to be located close to the boundary of the light cylinder radius, the distance at which the co-rotation velocity is equal to the speed of light. Interactions between charged particles (electrons \& positrons) and magnetic fields at the light cylinder generate pulsed, polarized emission. The physical mechanism behind the pulsed emission is still an open question --- a variety of models exist, each with different predictions for the pulse shape and polarization. Thus, polarization measurements of the pulsar can help constrain emission models.

The Crab is one of the best-studied objects in astrophysics. It is located $\sim2\:$kpc from Earth \citep{Trimble1973}, and with nebular luminosity $L\sim 1.3\times 10^{38}\ \mathrm{erg\ s}^{-1}$ \citep{Hester2008}, is bright enough to allow study of its morphology in great spatial detail with high statistical precision. At its center, the $P=33.6\:\mathrm{ms}$ PSR J0534+2200 is surrounded by a synchrotron nebula G184.6-5.8 that radiates strongly from radio to gamma-ray energies \citep{Ansoldi2016}. The inner nebula has polar jets and an equatorial torus wrapped around the termination shock that are prominent in X-rays. It is encased in a bubble-like structure of optical filaments formed by SN ejecta carving out cavities in the ambient ISM, with ``finger-like" protrusions extending inward from the filaments toward the lower-density synchrotron nebula. Diffuse radio emission exists throughout the nebula \citep{Hester2008}.

Polarization has been detected in both G184.6-5.8 and PSR J0534+2200. The first polarization measurements were made in 1954 at optical energies by two independent researchers \citep{Vashakidze1954, Dombrovsky1954}; high polarization levels confirmed the synchrotron origin of the nebular radiation, as suggested by \citet{Shklovsky1953}. This was the first identification of synchrotron radiation in any astrophysical source. Radio polarization measurements soon followed \citep{Mayer1957}, and later, with the advent of photon scattering polarimeters, soft X-rays \citep{Novick1972, Weisskopf1978}, hard X-rays \citep{Chauvin2017}, and gamma-rays \citep{Dean2008}. In each of these cases, the electric polarization angle integrated across the nebula was approximately along the torus symmetry axis (${\sim}125^\circ$, East of North), implying an azimuthal magnetic field \citep{Buhler2014}. 

Technological advances also later enabled temporal optical polarization studies, which allowed for phase-resolved analysis and separation of the pulsar and nebula components through isolation of the off-phase emission. In one of the most detailed optical studies of the Crab pulsar, \citet{Slowikowska2009} found that the polarization angle (PA) has a rapid monotomic sweep of about $+130^\circ$ through the main pulse (MP) and $+100^\circ$ through the interpulse (IP) and that the polarization degree (PD) seems to increase to a maximum before each pulse, then rapidly fall to a minimum close to the peak intensity. High angular resolution nebula polarization studies have also been conducted at optical energies \citep{Moran2013} and reveal high polarization levels in the inner knot and wisps (${\sim}60\%$ and ${\sim}40\%$, respectively) with directions oriented close to the pulsar spin axis. Similar studies at higher energies (e.g. X-rays) can provide critical information about the radiation of electrons and positrons closer to their injection site. 

The Imaging X-Ray Polarimetry Explorer (IXPE) \citep{Weisskopf2022}, the first space observatory dedicated to measuring X-ray polarization, has enabled that in the soft X-rays. With a nominal $2\,{-}\,8$ keV energy range, ${<}\,100\ \mu\mathrm{s}$ resolution, and ${<}\,30''$ HPD (half-power diameter), IXPE has detected polarization in several PWNe, and even a few phase bins near the pulse peak for the brightest pulsars \citep{Xie2022, Romani2023, Xie2024}, including the Crab \citep{Bucciantini2023}.

In 2022, IXPE observed the Crab for 90 ks. \citet{Bucciantini2023} detected PD $= 15.4\%$ and PA $= 105^\circ$ in the MP after ``off-pulse'' (OP) subtraction (the interval between the end of the interpulse and the start of the main pulse when the pulsar flux is at minimum). They also measured PD $= 24.1\%$ and PA $= 133^\circ$ in the OP. The phase-integrated polarization map revealed a toroidal magnetic field wrapped around the pulsar and asymmetric PD that does not align with the intensity map. Using an improved method to separate the pulsar and the nebula polarized fluxes, \citet{Wong2023} measured polarization in two MP phase bins (PD $= 9\%$, PA $= 97^\circ$ followed by PD $= 15\%$, PA $= 103^\circ$) and one IP phase bin (PD $= 16\%$, PA $= 141^\circ$). There was suggestion of a PA sweep through the MP, but more measurements were necessary to establish a clear pattern. They also extracted a polarization map for the pulsar-cleaned nebula, finding a toroidal magnetic field and PD asymmetries similar to those found in \citet{Bucciantini2023}. \cite{Mizuno2023} analyzed the nebula's magnetic field structure, including a comparison with a polarization model, and deferred spectropolarimetric analysis for future studies due to uncertainties in the spectral response at the time.

In 2023, IXPE observed the Crab for an additional 210 ks. In this paper, we analyze the full 300 ks IXPE dataset of the Crab, which yields ${\sim}\:$1.8$\times$ boost in signal-to-noise (S/N) and reduced systematic uncertainty relative to the initial discovery paper. Section \ref{sec:obs} describes the IXPE observations and the Chandra image used in the data analysis. Section \ref{sec:spec} presents the XSPEC spectral analysis of the nebula and its sub-regions. Section \ref{sec:pol} summarizes the reduction process for polarization, utilizing the simultaneous fitting technique of \citet{Wong2023}, and presents the nebular polarization map and the phase-varying pulsar polarization. Section \ref{sec:discussion} discusses possible physical interpretations of the spectral and polarization measurements.


\section{Observations} \label{sec:obs}
IXPE observed the Crab at three different epochs with a total ontime of about 300 ks: (1) February 21, 2022 $-$ March 8, 2022; (2) February 22, 2023 $-$ April 3, 2023; and (3) October 9 $-$ 10, 2023. Among all observations, the average livetime$\,:\,$ontime ratio was $0.923$. Associated with this campaign, the Chandra X-Ray Observatory (CXO) observed the Crab for an effective 1.33 ks exposure on March 15, 2022, one week after the first IXPE exposure.

\begin{figure}
    \hspace*{-1cm}
    \centering
    \includegraphics[width=1.2\linewidth]{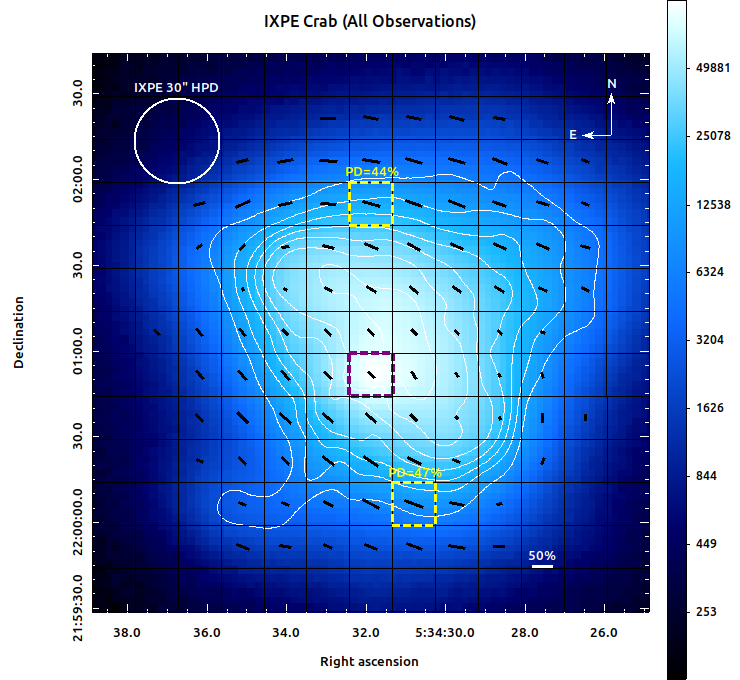}
    \caption{IXPE Crab count map ($3.25'\times3.25'$) displayed in log scale combining all three data epochs. Same white contours as in Figure \ref{fig:cxo} (right panel) to demonstrate the effect of IXPE ${<}\:30''$ HPD on the flux distribution. Fine nebular features in the CXO image (like the inner ring) are not visible, although the outer torus and southern jet can be discerned. Black boxes show the $15''$ grid used in the simultaneous fitting analysis. Black lines (perpendicular to the IXPE X-ray EVPA) reveal the toroidal magnetic field. Their lengths represent polarization degree with $\mathrm{PD}=50\%$ shown for reference. Only pixels with $>5\sigma$ detection and more than 20000 counts have polarization bars plotted. PSR J0534+2200 is located at the center of purple dotted box. High PD regions in the north and south are labeled. }
    \label{fig:ixpe}
\end{figure}

\subsection{IXPE} \label{subsec:ixpe_obs}
The first IXPE Crab observation (ObsID 01001099) was conducted in two segments, the first segment from February 21 $-$ 22, 2022 with a spacecraft roll angle of $158.0^\circ$ (East of North) and the second segment from March 7 $-$ 8, 2022 with a roll angle of $158.3^\circ$. The ontime for each segment was ${\sim}\:43$ ks and ${\sim}\:49$ ks, respectively. Because the offset between the optical axis and the spacecraft axes had not yet been measured and could not be taken into account during the target pointing, the optical axis was displaced from the target by ${\sim}\:2.74'$. We used \texttt{CIAO} tool \texttt{ixpecalcarf} to generate effective area and modulation response functions that account for this offset using the latest on-axis response files in the HEASARC CALDB database (XRT version 20231201, GPD version 20240125).

The second IXPE Crab observation (ObsID 02001099) was conducted in two segments, the first segment from February 22 $-$ 23, 2023 with a roll angle of $158.0^\circ$ and the second segment from April 1 $-$ 3, 2023 with a roll angle of $158.9^\circ$. Each segment ontime was ${\sim}\:74$ ks.

The third IXPE Crab observation (ObsID 02006001) was conducted in one segment from October 9 $-$ 10, 2023 with a roll angle of $339.0^\circ$ and ${\sim}\:60$ ks ontime.

For the second and third IXPE observations, we made vignetting-corrected response functions using \texttt{ixpecalcarf}. This was necessary to obtain a good agreement in the fitted spectral flux (see Section \ref{sec:spec}) between the observations with $2-3\%$ residual differences. 

We obtained the Level 2 data files for these observations from the HEASARC archive\footnote{\url{https://heasarc.gsfc.nasa.gov/docs/ixpe/archive}}. All data files were processed with the following steps before data analysis: (1) Particle and instrumental background events were removed according to the \citet{DiMarco2023} algorithm and by filtering for the good time intervals (GTI). The Di Marco background rejection algorithm removed less than $2\%$ of the total events. The GTI filter, which excludes observational periods with poor aspect \citep{Baldini2022}, removed less than $1\%$ of events from each detector unit (DU). (2) Barycentric correction was performed using the \texttt{barycorr} tool in \texttt{HEASoft V6.30.1}. The JPL-DE430 solar ephemeris was utilized with the position of the source set at R.A. = $5^\mathrm{h} 34^\mathrm{m} 31.86^\mathrm{s}$ and Decl. = $22^\circ00'51.3''$ (J2000). (3) The WCS (world coordinate system) was bore-sighted by comparing the $125'' \times 125''$ Stokes I data map centered on the pulsar in the MP window ($\Delta = 0.963-0.987$) with the simulated Stokes I map and adjusting each obesrvation's R.A. and Decl. in $0.1''$ increments to minimize the $\chi^2$-value. See Section \ref{sec:pol} for a description of the simulation procedure. By using the MP window, the alignment is keyed to the pulsar position, set at the aforementioned R.A. and Decl. This aspect correction, while small (typically $\sim1^{\prime\prime}$, always $<4^{\prime\prime}$), proved quite important to polarization measurements, improving the agreement of the IXPE data with our flux model by a factor of ${\sim}\:3.5\times$. (4) The pulse profile was folded with the \texttt{ixpeobssim} software \texttt{xpphase} tool. Table \ref{tab:ephemeris} lists the Jodrell Bank Observatory (JBO) ephemeris used for each observation epoch. Figure \ref{fig:ixpe} presents the total IXPE count map of the Crab, cropped to the $3.25' \times 3.25'$ region used for polarization analysis.

\subsection{Chandra} \label{subsec:cxo_obs}
CXO ObsID 23539 was obtained on March 15, 2022, one week after the conclusion of the first IXPE Crab observation. It was taken in 1:16 subframe mode (0.2s frame time) for 10 ks for a total livetime of 1.33 ks. These data were used to simulate the IXPE observation of the Crab PWN by passing the Level 2 event file through the IXPE instrument response using the \texttt{ixpeobssim V31.0.1} tool \texttt{xpobssim} and instantiating an \texttt{xChandraObservation} object. See Section \ref{sec:pol} for more details about the simulation procedure. Our IXPE observations extend out ${\sim}\:2.5$ years from the CXO observation. The bright inner wisps vary on the year-timescale, but the expected shifts are too small to affect the intensity on the IXPE PSF scale. Somewhat larger shifts are associated with the southern jet, but they appear on the decade-timescale so our CXO reference should be adequate.

Several artifacts needed to be removed from the Level 2 event file to produce a good quality file for use. \citet{Wong2023} noted two artifacts -- CCD saturation at the pulsar position and readout streaks due to out-of-time events from the pulsar and the nebula --- and reported correction methods. Here, we describe an improved technique to remove the nebula readout streak that reduces artificial jumps due to sampling in discrete regions. A $495''\times345''$ rectangle tilted along the readout direction encompassing the nebula readout streak was divided into a grid of $15''\times15''$ pixels. For each row, we estimated the excess counts and subtracted it from each pixel in that row. By correcting in smaller regions and adjusting the number of excess counts for each row, we produce a smoother readout-corrected image with fewer trail artifacts. 

We also found that pileup, which occurs when two or more events land on a CCD pixel within the same readout frame, was present in our CXO observation. The \texttt{CIAO} tool \texttt{pileup\_map} reported pileup fractions as high as ${\sim}\:20\%$ in the bright Doppler-boosted NW region of the nebula. Pileup effects underestimate the local count rate and distort the spectrum. To correct for this, we re-normalized the number of counts by $1/(1-p_f)^{1.5}$, where $p_f$ is the reported pileup fraction for a given pixel. The $1.5$ exponent heuristically addresses the effect of spectral distortions to the count rate. Note that this renormalization scheme only corrects the count rate and does not fix the spectrum. To do so, we would need to run a forward model of the pileup distortion with a template of the true spectrum of the nebula, which was not available. To minimize the effect of this spectral distortion on the polarization measurements, we have used a single $2-8$ keV energy band in our analysis. See Figure \ref{fig:cxo} for images before and after correcting for these artifacts as well as a distribution map of the pileup fraction across the nebula. 

\begin{table*}
\begin{center}
\hspace*{-1.6cm}
\begin{tabular}{ccccccc}
\hline
region & absorption & index & absorbed flux (2--8~keV) & PD  & PA &$\chi^{2}$/DoF\\
& $10^{22}~\mathrm{cm^{-2}}$& & $10^{-9}~\mathrm{erg~s^{-1}~cm^{-2}}$ & &deg &\\
\hline \hline
region~1 & $0.26\pm0.02$ & $2.160\pm0.007$ & 4.50 & $0.239\pm0.003$ & $137.8\pm0.3$ & 265.7/202\\
region~2N &0.30 (fix) & $2.234\pm0.003$ & 1.24 & $0.214\pm0.003$ & $152.1\pm0.4$ & 322.6/203\\
region~2E &0.30 (fix) & $2.233\pm0.004$ & 0.65 & $0.132\pm0.004$ & $134.6\pm0.8$ & 241.1/203\\
region~2S &0.30 (fix) & $2.303\pm0.005$ & 0.34 & $0.371\pm0.005$ & $150.8\pm0.4$ & 236.3/199\\
region~2W &0.30 (fix) & $2.164\pm0.003$ & 0.91 & $0.116\pm0.003$ & $135.7\pm0.8$ & 254.0/203\\
region~3N &0.30 (fix) & $2.323\pm0.003$ & 0.77 & $0.257\pm0.004$ & $157.0\pm0.4$ & 282.6/203\\
region~3E &0.30 (fix) & $2.350\pm0.005$ & 0.31 & $0.098\pm0.006$ & $128.0\pm1.6$ & 207.5/198 \\
region~3S &0.30 (fix) & $2.375\pm0.007$ & 0.16 & $0.363\pm0.008$ & $152.0\pm0.6$ & 214.4/192\\
region~3W &0.30 (fix) & $2.199\pm0.004$ & 0.57 & $0.088\pm0.004$ & $125.0\pm1.4$ & 264.1/203\\
\hline
\end{tabular}
\caption{\label{tab:nebula_spec}Summary of the Spectral Fits of Crab PWN}
\end{center}
\end{table*}

\begin{figure}
    \centering
    \includegraphics[width=\linewidth]{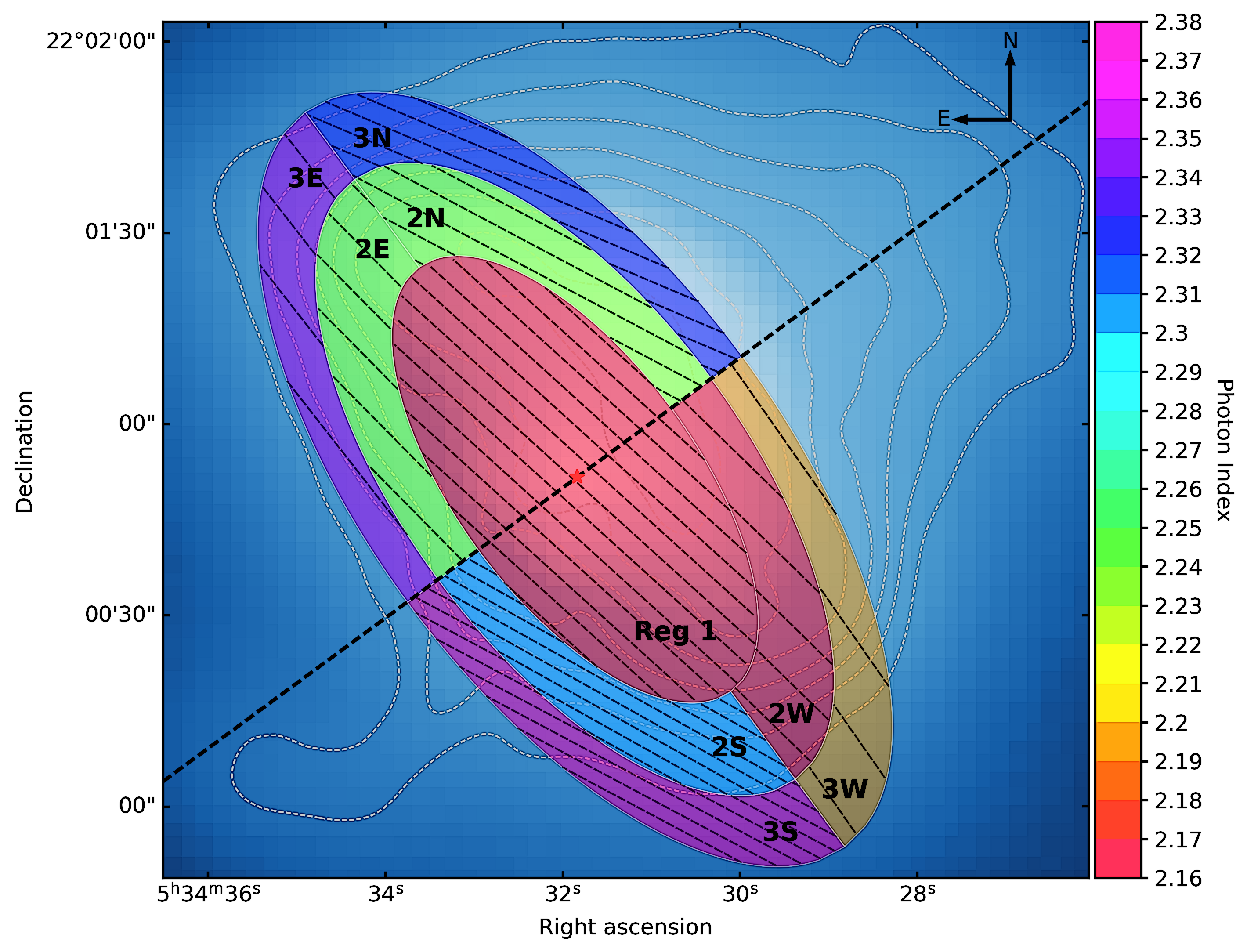}
    \caption{Spectropolarimetric analysis of selected regions of the Crab Nebula labeled according to Table \ref{tab:nebula_spec}. Shading represents photon index. Crosshatched lines represent magnetic field direction with line density $=13''/\mathrm{PD}$. Background blue heatmap overlaid with white contours (same as from Figure \ref{fig:ixpe}, right panel) represent IXPE count distribution. Dashed black line is the torus symmetry axis. Red star indicates the location of the pulsar.}
    \label{fig:nebula_spec}
\end{figure}

\section{XSPEC Spectral Analysis} \label{sec:spec}

We performed \texttt{XSPEC} spectropolarimetric analysis to investigate the positional dependence of spectral and polarization properties in the nebula. We defined a small ellipse, hereafter called ``Region~1," that contains the central X-ray torus with major and minor axes of $41.3''$ and $18.8''$ and a position angle of $126.3^\circ$ \citep{Ng2004}. We defined two outer regions, Region 2 and Region 3, with the same area as Region~1 and with respective inner and outer radii of $26.6''$, $58.4''$ and $32.6''$, $71.5''$. Both regions were further divided along the major and minor axes into four sub-regions, referred to as ``Region~2N", ``Region~3N", etc. See Figure \ref{fig:nebula_spec} for a diagram of the selected regions.

We extracted the Stokes I, Q, and U spectrum of each region using the standard software \texttt{ixpeobssim V31.0.1}. For Region~1, we used only data in the off-pulse period ($\Delta = 1.563-1.863$) to minimize pulsar contamination. For the other regions, the spectral and polarimetric parameters change by less than 5\% if we use all data so we do not apply a phase cut to simplify the analysis. For background subtraction, we extracted a spectrum from an annulus centered on the pulsar position with the inner and outer radius of $2.5'$ and $3.0'$, respectively. We found that the background has negligible effects for all regions and do not subtract it in the spectral analysis for simplicity. As first discussed by \citet{Bucciantini2023Leakage}, we also needed to address the effect of ``leakage", which is the spatial spreading of polarized flux, preferentially in the direction of the polarization, due to imperfect reconstruction of the photon position in the IXPE detector. \citet{Dinsmore2024} present a correction technique using detailed 2D sky-calibrated IXPE PSFs with a code library publicly available in the Github repository \texttt{leakagelib}\footnote{\url{https://github.com/jtdinsmore/leakagelib}}. Using the code and the polarization parameters obtained through simultaneous fitting (see Section \ref{sec:pol}), we generated 5$''$ Stokes I, Q, and U leakage maps binned into 4 energy bands ($2-2.8$, $2.8-4$, $4-5.6$, and $5.6-8$~keV). For each region, we extracted the Stokes I, Q, and U leakage spectra, normalized by the Stokes I spectrum, and fit them with a power-law model. Then we calculated the model leakage spectrum in the standard energy binning (0.04 keV) and subtracted it from the total spectrum. 

\begin{figure}
    \centering
    \includegraphics[width=\linewidth]{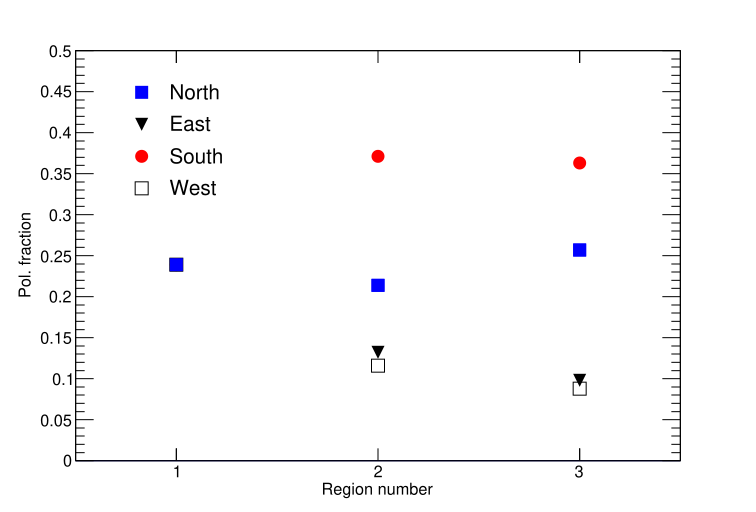}
    \includegraphics[width=\linewidth]{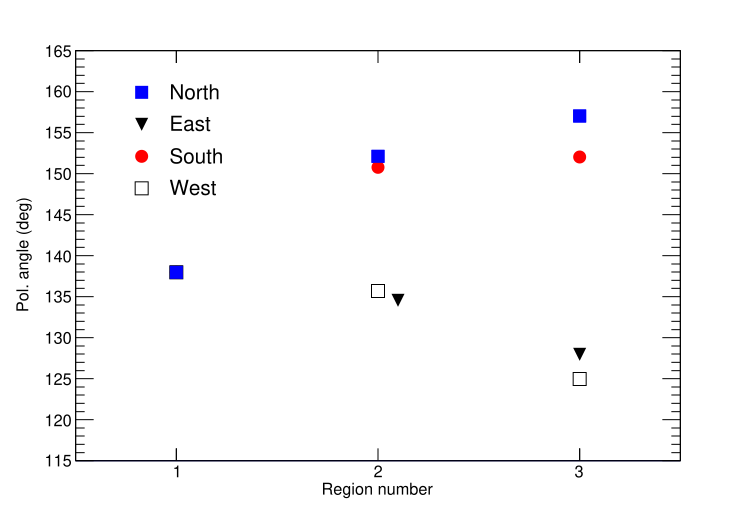}
    \includegraphics[width=\linewidth]{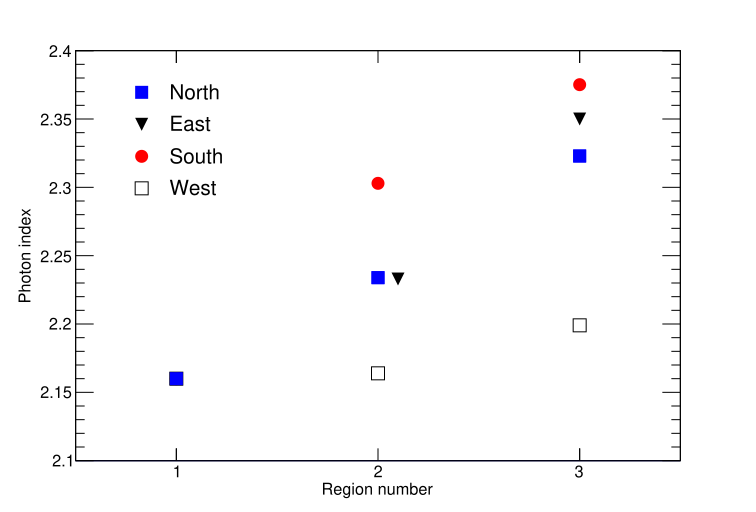}
    \caption{Spectropolarimetric fits for different regions of the Crab Nebula. Error bars are small, comparable to the marker size, and not presented. Region~2E has been shifted horizontally in middle and bottom panels for clarity. Polarization degree and angle diverge between north/south and east/west regions, as expected from a toroidal magnetic field. Photon index increases with radius, as might be anticipated due to synchrotron burnoff. It is also asymmetric relative to the torus symmetry axis, with the hardest emission in the West with the lowest polarization degree and the softest emission in the South with the highest polarization degree.}
    \label{fig:FitResult}
\end{figure}

Using our leakage-corrected Stokes I, Q, and U spectra, we performed spectropolarimetric fits using the \texttt{xspec} command in \texttt{HEASoft V6.32.1} with an absorbed power-law model and energy-independent polarization (\texttt{tbabs} $\times$ \texttt{powerlaw} $\times$ \texttt{polconst} in \texttt{xspec}). To mitigate the low photon statistics, we fixed the absorption density to the canonical value of $0.3 \times 10^{22}~{\rm cm^{-2}}$ \citep{Mori2004} for regions other than Region~1. We fitted the Region 1 absorption density to obtain better fit statistics, but the fitted value remained close (within $2\sigma$) of the canonical value. Fit results are summarized in Table \ref{tab:nebula_spec} (with 1$\sigma$ statistical errors) and Figures \ref{fig:nebula_spec} and \ref{fig:FitResult}.

We were able to obtain agreement of the total flux within $2-3\%$ between observations. Our photon indices are systematically larger by a few percent than those reported in the literature. For example, we find a photon index of $2.160\pm 0.007$ for the central region but \citet{Mori2004} reports values around ${\sim}\,2.0$ for the inner nebula. This difference is likely due to contributions from the softer, outer regions of the nebula \citep{Weisskopf2000} and possibly also instrument calibration. Effects on the polarization parameters and the relative value of the spectral parameters should be much smaller.

\begin{figure}
    \centering
    \includegraphics[width=\linewidth]{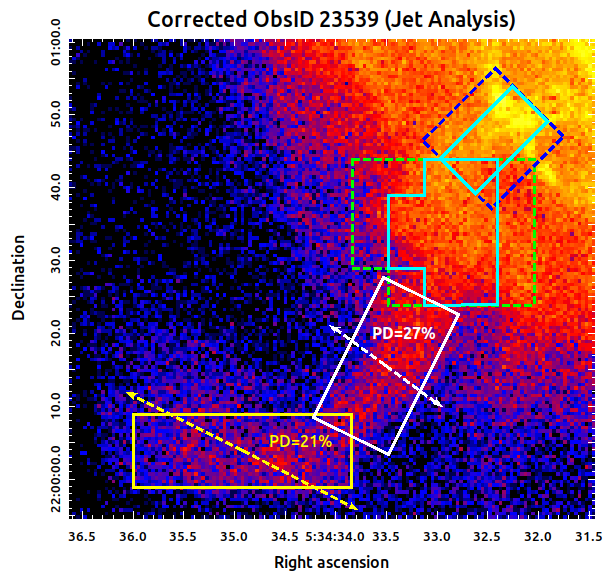}
    \caption{Analysis of the Crab Nebula jet polarization. Jet was divided into the yellow (tip), white (body), and cyan (base) regions. Green and blue dashed boxes represent the flanking regions used for background subtraction. No conclusive measurements in the base were made. In the body and tip, polarization degree is labeled and magnetic field direction (perpendicular to the polarization angle) is indicated by the dashed arrows. CXO ObsID 23539 displayed in the background to visualize location of the regions along the jet.}
    \label{fig:nebula_jet}
\end{figure}

\begin{figure*}
    \centering
    \includegraphics[width=\linewidth]{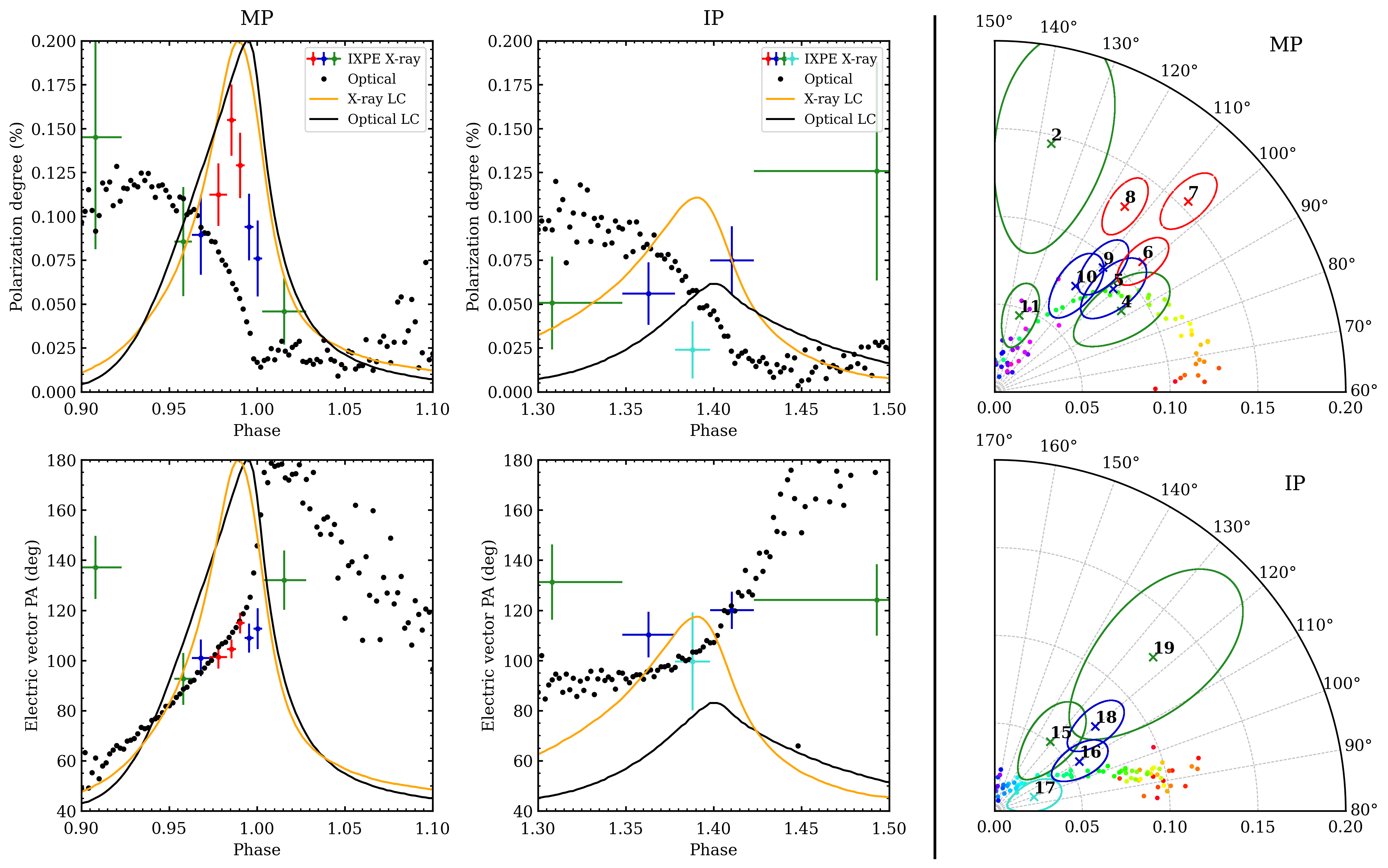}
    \caption{Comparison of X-ray and optical \citep{Slowikowska2009} PD and PA polarization measurements of the Crab Pulsar. Only MP ($\Delta \mathrm{P1} = 0.9-1.1$) and IP ($\Delta \mathrm{P2} = 1.3-1.5$) are displayed. Red points are ${>}\,5\sigma$ significance, blue ${>}\,3\sigma$, and green ${>}\,1.9\sigma$. One turquoise ${<}\,1.9\sigma$ bin in the IP included for continuity. PD and PA versus phase are shown in left and middle columns. Normalized light curves for X-ray (orange) and optical (black) are included for reference. Polar representations are shown in the rightmost column, where X-ray measurements with $1\sigma$ contours and phase numbering (starting from 1) are plotted against optical values colored from red to blue to show progression through the pulse. Positive ${\sim}\,40^\circ$ X-ray PA sweep can be clearly seen through the MP. See Section \ref{subsec:pulsar_pol} for detailed discussion.}
    \label{fig:pulsar_pdpa}
\end{figure*}

\section{Simultaneous Fit Polarization} \label{sec:pol}

\begin{figure*}
    \centering
    \includegraphics[width=\textwidth]{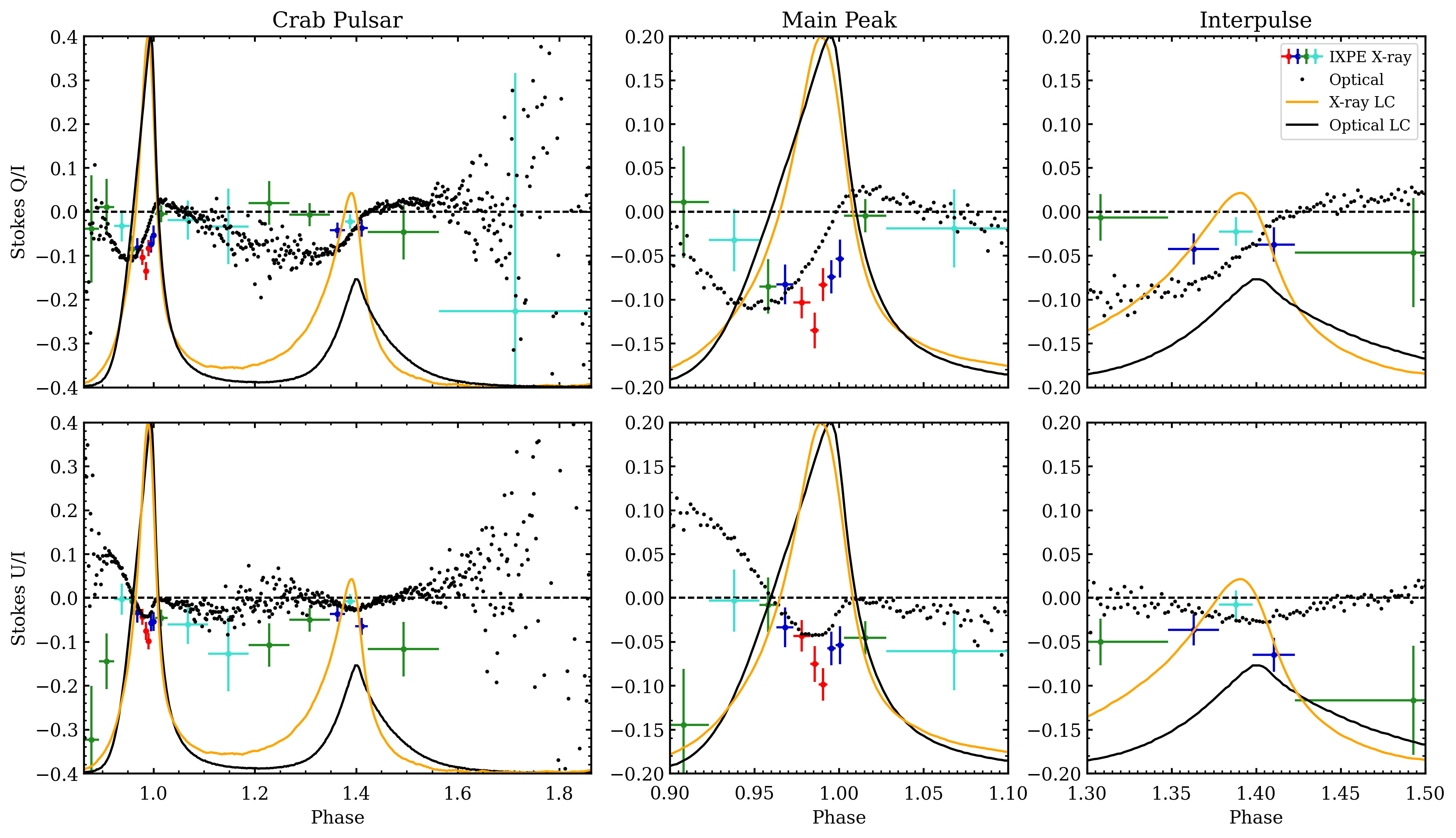}
    \caption{Comparison of X-ray and optical \citep{Slowikowska2009} Stokes q and u polarization measurements of the Crab Pulsar. Full phase profile (left), main pulse (middle, $\Delta \mathrm{P1} = 0.9-1.1$), and interpulse (right, $\Delta \mathrm{P2} = 1.3-1.5$) are depicted. Red points are ${>}\,5\sigma$ significance, blue ${>}\,3\sigma$, green ${>}\,1.9\sigma$, and turquoise ${<}\,1.9\sigma$. Normalized light curves for X-ray (orange) and optical (black) are included for reference. ${>}\,1\sigma$ differences can be seen between the two energies, see Section \ref{subsec:pulsar_pol} for detailed discussion. Different radiation processes or sites may be responsible for the differences in the polarization at the two energies.}
    \label{fig:pulsar_xopt}
\end{figure*}

We used the simultaneous fitting technique described in \citet{Wong2023} to extract the nebula and the pulsar polarization parameters. The technique requires a model for the pulsar and the nebula flux as a function of phase, energy, and spatial position. The Crab pulsar model was created by running an \texttt{ixpeobssim} simulation of a periodic point source with phase-varying spectral parameters obtained from CXO HRC-LETGS data by \citet{Weisskopf2011}. The Crab nebula model was made by instantiating an \texttt{xChandraObservation} object with the artifact-cleaned CXO ObsID 23539 in \texttt{ixpeobssim}. For both models, the different efficiencies of the CXO and IXPE detectors were accounted for by passing the CXO measurements through the ratio of the IXPE : CXO effective areas and applying the IXPE point spread function. A 750 ks simulation, more than $10\times$ longer than any one segment exposure, was generated for each segment using its specific telescope roll and instrument response function and normalized by its specific ontime duration. A small scaling factor was applied to match the simulated and the observed IXPE light curves. The pulsar scaling factor was ${\sim}\,0.75$ and the nebula scaling was ${\sim}\,0.90$. A small normalization constant, fixed at 1 for DU1 and within 3.5\% for the other DUs, was applied for each detector unit to account for calibration differences.

Note that the \citet{Weisskopf2011} Crab pulsar spectrum was derived using $0.3-3.8$ keV CXO data within a radius of $1.63''$. The spectral analysis done by \citet{Massaro2000} with BeppoSAX MECS used $1.6-10$ keV data, closer to the IXPE nominal energy range, but with an extraction radius of $4'$, which has considerably more nebula contribution. They measured phase-varying photon indices about $+0.2$ larger than those found by \citet{Weisskopf2011}. To test the sensitivity of our method to the spectral model, we ran the same analysis using a softer pulsar spectrum with photon indices boosted by $+0.2$ and found that the Stokes q and u parameters differed by at most ${\sim} 1.5\,\sigma_{\mathrm{error}}$ of the original fit for the pulsar and $0.3\,\sigma_{\mathrm{error}}$ for the nebula. \citet{Weisskopf2011} isolates the pulsar more reliably than \citet{Massaro2000} due to the use of a smaller extraction region so the larger photon index measured by \citet{Massaro2000} should at least partially be attributed to nebula contamination. Therefore, these values represent upper limits for the uncertainty of our measurements due to the imperfect estimate of the pulsar photon index.

To correct for ``leakage" effects, we took an iterative approach to estimate and subtract the leakage from the data and find the leakage-corrected polarization parameters: (1) an initial fit for the polarization is performed using uncorrected IXPE data; (2) these parameters are input into \texttt{leakagelib} to calculate Stokes I, Q, and U leakage, which we subtract from the data; (3) we fit for new polarization parameters using the leakage-corrected dataset; and (4) this process is repeated until the average fractional change of the parameters is less than $10^{-5}$, a standard value used in (i.e. Python) software packages for relative comparisons. We found that about three iterations were required to reach convergence and that the Stokes q and u parameters changed more substantially (within about $\pm 0.03$) for the nebula; for the pulsar, most (95\%) of the bins changed by within $\pm 0.01$.

As described in \citet{Wong2023}, simultaneous fitting is a binned analysis method. For our analysis, we used 20 variable-width phase bins, tailored to be narrower in the pulses to probe the rapid sweep of the pulsar polarization at these phases, and a $13\times 13$ $15''$ grid of spatial pixels. See Table \ref{tab:pulsar_pol} for exact phase bin selection in the main and inter-pulses. Since the leakage correction algorithm requires fine (max $5''$) spatial bins to resolve the PSF structure, we calculated the leakage with $5''$ pixels and regrouped them into $15''$ pixels before subtracting it from the binned data.

For energy binning, we used a single $2-8$ keV bin because finer energy binning significantly increases the number of low-count (${<}\,10$ counts) bins. Our fitting approach solves for the polarization parameters using least-squares regression, and thus, assumes Gaussian-distributed data. For $N$ events with the $\psi$-distribution function described in \citet{Kislat2015}, Stokes Q and U are essentially Gaussian by $N=10$. With the aforementioned phase and spatial binning and three equally-spaced energy bins within $2-8$ keV, 52\% of data bins would have fewer than 10 counts, about half from the $6-8$ keV energy bin. Without energy binning, 14\% of the bins exceed 10 counts. To use energy bins while minimizing low-count bins, we could use fewer phase bins or larger pixels, but this would degrade resolution of the rapid changes in polarization from the pulsar emission and/or over-smooth the PSF structure, which separates the pulsar from the nebula flux. The nebula is indeed softer with a photon index of ${\sim}\,2$ \citep{Willingale2001, Mori2004} compared to ${\sim}\,1.4-1.7$ \citep{Weisskopf2011} for the pulsar, so an energy-dependent analysis could make a slight improvement in separating the pulsed and the nebular signals. However, this would require substantial additional exposure or a full-Poisson statistics likelihood analysis (which would be much more computationally expensive than our least squares solution), and so we have elected to retain the best spatial and temporal resolution in a single energy bin. 

In summary, we have utilized the simultaneous fitting technique to separate the nebula and the pulsar polarization parameters. This technique requires Stokes I models, which we generated by taking the nebula spectral map and the pulsar phase-varying spectrum obtained from CXO observations and passing them through the IXPE instrument response. The model and the data were binned into 20 variable-width phase bins and $13\times 13$ $15''$ pixels and an energy range of $2-8$ keV. Leakage removal was performed in conjunction with the simultaneous fitting in an iterative process using the \texttt{leakagelib} code. The fit statistic of our final iteration was $\chi_{\mathrm{red}}^2=1.23$ with $\chi^2=124066.0$ and $\mathrm{DOF}=100708$. The $\chi_{\mathrm{red}}^2 > 1$ may possibly be due to remaining mismatches between the observation and our flux model. Our model can be further refined with improved calibration of the instrument response functions, correction of pileup spectral effects in the CXO image, energy-dependent PSFs, and joint IXPE/CXO observations to obtain the most current state of the nebula.

\begin{table*}
\begin{center}
\hspace*{-3.7cm}
\begin{tabular}{  c | c | c | c | c | c | c } 
\hline
\multicolumn{7}{c}{\bf{Simultaneous Fit Polarization Measurements of the Crab Pulsar}} \\
 \hline
 \hline
    & & Phase & q & u & PD (\%) & PA ($^\circ$) \\
 \hline
 \hline
 \multirow{9}{*}{P1} & $2^{\dag}$ & $0.893-0.923$ & $0.0108 \pm 0.0638$ & $-0.1446 \pm 0.0637$ & $14 \pm 6$ & $137 \pm 13$ \\
 \cline{2-7}
 \multirow{8}{1.0cm}{} & $4^{\dag}$ & $0.953-0.963$ & $-0.0852 \pm 0.0311$ & $-0.0079 \pm 0.0311$ & $9 \pm 3$ & $93 \pm 10$ \\
 \cline{2-7}
 \multirow{7}{1.0cm}{} & 5 & $0.963-0.973$ & $-0.0829 \pm 0.0228$ & $-0.0336 \pm 0.0228$ & $9 \pm 2$ & $101 \pm 7$ \\
 \cline{2-7}
 \multirow{6}{1.0cm}{} & 6 & $0.973-0.983$ & $-0.1036 \pm 0.0179$ & $-0.0433 \pm 0.0179$ & $11 \pm 2$ & $101 \pm 5$ \\
 \cline{2-7}
 \multirow{5}{1.0cm}{} & 7 & $0.983-0.988$ & $-0.1352 \pm 0.0203$ & $-0.0753 \pm 0.0203$ & $15 \pm 2$ & $105 \pm 4$ \\
 \cline{2-7}
 \multirow{4}{1.0cm}{} & 8 & $0.988-0.993$ & $-0.0831 \pm 0.0186$ & $-0.0987 \pm 0.0186$ & $13 \pm 2$ & $115 \pm 4$ \\
 \cline{2-7}
 \multirow{3}{1.0cm}{} & 9 & $0.993-0.998$ & $-0.0741 \pm 0.0190$ & $-0.0577 \pm 0.0190$ & $9 \pm 2$ & $109 \pm 6$ \\
 \cline{2-7}
 \multirow{2}{1.0cm}{} & 10 & $0.998-1.003$ & $-0.0534 \pm 0.0216$ & $-0.0539 \pm 0.0216$ & $8 \pm 2$ & $113 \pm 8$ \\
 \cline{2-7}
 \multirow{1}{1.0cm}{} & $11^{\dag}$ & $1.003-1.028$ & $-0.0047 \pm 0.0190$ & $-0.0455 \pm 0.0190$ & $5 \pm 2$ & $132 \pm 12$ \\
 \hline
 \hline
 \multirow{5}{1.0cm}{P2} & $15^{\dag}$ & $1.268-1.348$ & $-0.0066 \pm 0.0265$ & $-0.0502 \pm 0.0265$ & $5 \pm 3$ & $131 \pm 15$ \\
 \cline{2-7}
 \multirow{4}{1.0cm}{} & 16 & $1.348-1.378$ & $-0.0425 \pm 0.0179$ & $-0.0364 \pm 0.0179$ & $6 \pm 2$ & $110 \pm 9$ \\
 \cline{2-7}
 \multirow{3}{1.0cm}{} & $17^{*}$ & $1.378-1.398$ & $-0.0225 \pm 0.0163$ & $-0.0079 \pm 0.0163$ & $2 \pm 2$ & $100 \pm 20$ \\
 \cline{2-7}
 \multirow{2}{1.0cm}{} & 18 & $1.398-1.423$ & $-0.0374 \pm 0.0195$ & $-0.0649 \pm 0.0195$ & $7 \pm 2$ & $120 \pm 7$ \\
 \cline{2-7}
 \multirow{1}{1.0cm}{} & $19^{\dag}$ & $1.423-1.563$ & $-0.0466 \pm 0.0624$ & $-0.1168 \pm 0.0623$ & $13 \pm 6$ & $124 \pm 14$ \\
 \hline
\end{tabular}
\end{center}
\caption{\label{tab:pulsar_pol} Significant (${>}3\sigma$) and marginally significant (${>}1.9\sigma$, dagger) X-ray polarization measurements of the Crab Pulsar. One ${\sim}\,1.5\sigma$ P2 phase bin (starred) has been included for continuity. Refer to Figures \ref{fig:pulsar_pdpa} and \ref{fig:pulsar_xopt} for visual representation. Note that the marginally significant measurements have significant PD$-$PA covariance so the 1D errors reported here are not fully representative. (Stokes q and u are independent so their 1D errors are reliable.) Phases are numbered starting from 1.}
\end{table*}

\subsection{Crab Nebula} \label{subsec:nebula_pol}

The Crab Nebula polarization map is depicted in Figure \ref{fig:ixpe}. The black lines indicate the magnetic field direction, perpendicular to the measured electric vector polarization angle (EVPA), with lengths scaled by the polarization degree. $5\sigma$ significance cut and $20000$ count flux cut have been applied. The most polarized regions are located in the north and south edges of the torus, with the highest $\mathrm{PD} = (44 \pm 1)\%$ and $\mathrm{PD} = (47 \pm 1)\%$ in these regions, respectively. Two pixels in the jet have significant polarization: one in the body, where the magnetic field appears perpendicular to the jet, and another at the tip, where it appears parallel, with $\mathrm{PD} = (22 \pm 2)\%$ and $\mathrm{PD} = (19 \pm 2)\%$, respectively. To investigate the jet polarization further, we performed simultaneous fitting with $25\times25$ $5''$ pixels (and 16 phase bins, to minimize the number of low-count bins to ${\sim}\:10\%$) to obtain a higher resolution map, divided the map into different regions along the jet, and determined the integrated polarization in each region. We verified that the $5''$ polarization map is consistent with that of the original binning, with the most polarized regions located in the same high-PD regions labeled in Figure \ref{fig:ixpe} and having similar polarization degree and polarization angle values throughout the nebula.

Figure \ref{fig:nebula_jet} depicts the result of the jet analysis. We created four regions along the jet: two cyan regions at the base (with bracketing blue and green regions for background subtraction) and one region each for the body (white) and the tip (yellow). For the tip and body, no background subtraction was necessary since the torus does not overlap. We simply summed the pixels within each region and found that the body is polarized with $\mathrm{PD}=(27 \pm 1)\%$ and $\mathrm{PA} =(144 \pm 1)^\circ$ and the tip is polarized with $\mathrm{PD}=(21 \pm 2)\%$ and $\mathrm{PA}=(153 \pm 3)^\circ$. As shown in Figure \ref{fig:nebula_jet}, these polarization angles suggest that the magnetic field is oriented perpendicular and parallel relative to the jet axis in the body and the tip, respectively. Only $5''$ pixels with ${>}3\sigma$ polarization measurements were included in the calculation.

Some concern might be made about the potential for contamination from the torus and adjacent jet regions. Indeed, using \texttt{ixpeobssim} to simulate the central nebula region and the two jet regions individually, we estimate that the torus contributes approximately 23\% and 7\% of the total flux in the body and tip regions, respectively, and that the body contributes ${\sim}\:13\%$ of the total flux in the tip, and that the tip contributes ${\sim}\:5\%$ of the total flux in the body. This means that $28\%$ and $20\%$ of the flux in the body and the tip, respectively, may be attributed to these regions. In fact, the total background fraction, including contributions from other areas of the nebula, is estimated to be ${\sim}\:60\%$ and ${\sim}\:50\%$ for the body and tip, respectively. To determine how significantly the background affects our measurements, we ran a simultaneous fitting procedure where we included the effect of the PSF flux redistribution in the nebula (it is always computed for the pulsar). That is, for each $5''$ pixel, we modeled its contribution to each of the other pixels in the flux map. Using this model should eliminate background contributions on greater than $5''$ scales. For this fine pixel scale, we needed additionally to regularize the cost function (the objective function that is minimized in the least-squares analysis) with a penalty for large swings between adjacent pixels. We obtained polarization values consistent with our standard analysis, with $\mathrm{PD}=30\%$ and $\mathrm{PA}=145^\circ$ in the body and $\mathrm{PD}=22\%$ and $\mathrm{PA}=159^\circ$ at the tip. This suggests that our polarization measurements are not significantly biased by the background. However, given the high background percentage, the true uncertainty may be larger than the simple statistical error reported here. Higher spatial resolution would be helpful to isolate the jet polarization more confidently.

We also attempted to measure the polarization in the base regions. Since the jet overlaps with the torus in these regions, we selected flanking fields to estimate the torus flux and subtracted it from the flux in the base. Our results were inconclusive, producing aphysical $\mathrm{PD}\,{>}\,100\%$ with large uncertainties. We will note that, in the $5''$ polarization map, the polarization degree of the pixels in the base regions were slightly lower than those of the pixels in the bracketing (background) regions, which suggests that the jet may have a different polarization orientation than the background torus. Higher angular resolution polarization imaging would be immensely helpful in isolating the jet to test this hypothesis.

\subsection{Crab Pulsar} \label{subsec:pulsar_pol}

The Crab pulsar X-ray polarization measurements are plotted against the optical (\citet{Slowikowska2009}, obtained via private communication) in traditional PD/PA format (Figure \ref{fig:pulsar_pdpa}), useful for model comparison, and in Stokes format (Figure \ref{fig:pulsar_xopt}). In the traditional plots, polarization parameters below $3\sigma$ cannot reliably be reported with 1D error bars due to $\mathrm{PD}-\mathrm{PA}$ covariance. Hence, we have restricted the phase range to the MP and the IP and marked marginally significant (and for continuity, one ${<}\,1.9\sigma$ in the IP) measurements with a different color. The polarization values and uncertainties are listed in Table \ref{tab:pulsar_pol}.

Among the 20 phase bins, we detect polarization in six phases within the MP and two phases within the IP. In the MP, the PD appears to rise from ${\sim}\,7.5\%$ and reach a maximum of ${\sim}\,15\%$ near the peak phase before falling back to its previous level. By comparison, in the optical, the PD is at a constant ${\sim}\,12.5\%$ at the rising edge of the pulse and falls to ${\sim}\,2.5\%$ right after the optical peak phase. In the X-rays, the PA has an approximately $+40^\circ$ sweep between phases 0.958 and 1.0155. Before the peak, the X-ray PA curve nearly matches the optical PA curve, but afterwards, it appears to rise more slowly and lag behind the optical curve. In the IP, the two significant measurements have PD between 5\% and 10\%, bracketing one low-significance measurement near the pulse peak. By comparison, the optical PD is at a constant ${\sim}\,10\%$ at the rising edge of the pulse and falls through the pulse to $2.5\%$. The X-ray IP PA values lie near the optical values and hint at an upward sweep.

Speculating that the low X-ray PD at the center of the IP could be attributed to a rapid angle sweep across this bin, we tried dividing it into two equally-spaced bins. However the polarization still could not be significantly measured in these smaller bins. In addition, we tested for a smooth polarization sweep across this phase range, by partitioning the data in small (0.00017-width) bins, subtracting the background nebula polarization using the simultaneous-fit measurements of Section \ref{subsec:nebula_pol}, and fitting a linear model. The polarization slope was only significant at ${\sim}\,1.2\sigma$. With this result, we are not able to conclude whether there is a PA sweep at the IP center.

In Figure \ref{fig:pulsar_xopt}, we can see that many bins have ${>}1\sigma$ differences between the optical and X-ray in the Stokes parameters. In the MP, the X-ray Stokes q values fall to a minimum at the center of the pulse before sweeping rapidly up while the optical counterpart monotomically increases through the pulse. Both the X-ray and optical Stokes u values dip leading up to the pulse peak, then rise back up, but the X-ray curve has a sharper dip and reaches a lower minimum. In the IP, no conclusive trends can be inferred with only two significant X-ray measurements. Notably, however, the X-ray and optical Stokes values do differ by ${\gtrsim}\:1\sigma$ in both IP bins.

\section{Discussion} \label{sec:discussion}

While detailed emission models are needed for a full confrontation of these data with theory, we can discuss qualitative implications of our polarization measurements for the physical conditions in the Crab PWN and pulsar. 

From our spectropolarimetric analysis, we find that the polarization angle deviates from that of Region~1 in diametric ways between the north/south and east/west outer regions, as expected in a toroidal geometry. Also, we can see a clear directional dependence of the photon index, suggesting that the energy of the emission is affected not only by synchrotron burnoff, which increases with the distance from the termination shock and was also observed by \citet{Weisskopf2000}, but also by some other effect. Although the root cause of this dependence is not clear, it is worth noting that photon index is hardest in the West, where the polarization degree is smallest, and is softest in South, where the polarization degree is largest. \citet{Mori2004} also report an extended wing of hard emission towards the southwestern edge of the nebula. This observation might be consistent with turbulent (re-)acceleration, which would lead to a harder spectrum in a local region of lower polarization. Indeed, Regions 2W and 3W sit adjacent to the western bay, impingement with which could cause increased turbulence.


In the nebula polarization map, the high polarization at the north and south of the torus, with $\mathrm{PD}\;{\sim}\;45-50\%$, and the depolarization on the northeast and southwest sides, where polarization direction changes rapidly, are consistent with the findings of \citet{Bucciantini2023}. In the jet, the polarization degree drops by ${\sim}\;6\%$ as one moves downstream along the jet and the angle changes relative to the jet axis from perpendicular to parallel. This observation may be explained by the growth of kink instabilities along the jet. In the 3D relativistic MHD simulations by \citet{Mignone2013}, the jet is subject to instabilities while freely propagating within the PWN flow. Such behavior is seen in CXO monitoring of the Vela pulsar jet \citep{Pavlov2003}. They find that the jet deflection radius increases with the magnetization parameter $\sigma$, suggesting that magnetic instabilities are driving the deflection. They also see that the horizontally-averaged direction of the magnetic field is initially perpendicular to the flow velocity but bends and acquires a parallel component by the tip. Our jet polarization measurements are consistent with these simulated observations. Alternatively, collision against a dense medium (e.g.~an optical filament) may be causing a hoop stress-confined jet (with initial field dominated by a toroidal component) to bend, compressing and amplifying the magnetic field parallel to the collision shock front. If accompanied by turbulence, this would also lower the polarization fraction, as seen here.

From our pulsar measurements, we obtain a 3$\sigma$ upper-limit of the bridge ($\Delta = 1.028-1.348$) phase-averaged $\mathrm{PD}=22\%$. Phase-averaged total pulsar polarization is insignificant with Stokes $\mathrm{q} = -0.007\pm0.047$ and Stokes $\mathrm{u} = -0.075\pm0.047$.

Comparing our polarization sweep with that previously measured by \citet{Wong2023} using only the 2022 IXPE Crab observation, a curious difference may be noted: they measured Stokes $\mathrm{u} = -0.158\ \pm\ 0.039 $ at the single phase bin under the IP peak, which is ${>}2\sigma$ higher than the IP Stokes u measurements reported here. To investigate this, we ran simultaneous fitting for each observation segment and found that Stokes u was approximately $-0.15$ for the first segment at the single bin under the IP peak (bin 17). Meanwhile, the other segments reported lower values within $\pm0.05$. The large Stokes u reported in \citet{Wong2023} is thus likely a statistical fluctuation and is not changed by the modeling improvements described in the present paper.

We have attempted to compare our pulsar polarization measurements with a simple analytic striped wind model \citep{delZanna2006, Petri2013}. From our analysis, we find that the pulse morphology is most sensitive to the assumed wind velocity structure due to the relativistic beaming of emission from the current sheet. If the emission region extends more than a few light-cylinder radii ($R_{lc}$) one finds narrow MP/IP peaks with similar intensity, incompatible with observations. While emission confined within a few $R_{lc}$ produces a plausible intensity profile, the polarization sweeps are quite difficult to reproduce in this simple model. Polarization is more sensitive to the assumed magnetic field geometry, and a simple split-monopole model appears to inadequately capture the observed behavior. We note that geometry-dependent reconnection could modulate the emissivity of the current sheet and turbulence growth across the emission zone may affect the polarization degree; these may be important factors to consider in future modeling.

While a detailed treatment goes beyond the scope of this paper, it is important to note that, while the excellent optical polarization data of \citet{Slowikowska2009} have been available for some time, no model has been produced which can match its behavior in detail. The fact that IXPE now detects polarization sweep structure at X-ray energies but with substantial differences with the optical values provides a new handle on the problem. Since the X-rays are generated by higher energy electrons, they have different weighting along the emission surface, and possibly different propagation and self-absorption effects, so they provide a new lever to probe this long-standing astrophysical puzzle. Of course, finer and more extended phase-resolved X-ray polarimetry would enhance this probe. In particular, additional IP phase bins are really needed to compare its structure to that seen in the optical. It could also be important to get a few well-measured bins in the bridge region, as this probes emission away from the caustic-dominated peaks. Such data may be attained with very deep IXPE observations or may require a future X-ray polarization mission with better spatial resolution and larger effective area.  
\bigskip

Acknowledgements:
This work was supported in part through NASA grant NNM17AA26C
administered by the Marshall Space Flight Center.

The Imaging X-ray Polarimetry Explorer (IXPE) is a joint US and Italian mission.  The US contribution is supported by the National Aeronautics and Space Administration (NASA) and led and managed by its Marshall Space Flight Center (MSFC), with industry partner Ball Aerospace (contract NNM15AA18C).  The Italian contribution is supported by the Italian Space Agency (Agenzia Spaziale Italiana, ASI) through contract ASI-OHBI-2022-13-I.0, agreements ASI-INAF-2022-19-HH.0 and ASI-INFN-2017.13-H0, and its Space Science Data Center (SSDC) with agreements ASI-INAF-2022-14-HH.0 and ASI-INFN 2021-43-HH.0, and by the Istituto Nazionale di Astrofisica (INAF) and the Istituto Nazionale di Fisica Nucleare (INFN) in Italy.  This research used data products provided by the IXPE Team (MSFC, SSDC, INAF, and INFN) and distributed with additional software tools by the High-Energy Astrophysics Science Archive Research Center (HEASARC), at NASA Goddard Space Flight Center (GSFC). N.B. was supported by the INAF MiniGrant ``PWNnumpol - Numerical Studies of Pulsar Wind Nebulae in The Light of IXPE." F.X. is supported by National Natural Science Foundation of China (Grant No. 12373041). T. M. was supported by JSPS KAKENHI Grant Number 23K25882. I.L. was supported by the NASA Postdoctoral Program at the Marshall Space Flight Center, administered by Oak Ridge Associated Universities under contract with NASA. This paper employs the Chandra dataset, obtained by the Chandra X-ray Observatory, contained in\dataset[doi: 10.25574/cdc.264]{https://doi.org/10.25574/cdc.264}.

\facilities{IXPE, CXO}

\software{\texttt{IXPEobssim V31.0.1} \citep{Baldini2022}, \texttt{leakagelib} \citep{Dinsmore2024, jack_dinsmore_2024_10483298}}

\bibliography{main}{}
\bibliographystyle{aasjournal}

\end{document}